\documentclass[preprint,showpacs,preprintnumbers,amsmath,amssymb,superscriptaddress,prl]{revtex4}
\usepackage{amsmath}

\usepackage{amssymb}

\usepackage[pdftex]{graphicx}

\usepackage{dcolumn}

\usepackage{bm}

\usepackage{color}

\def\be{\begin{eqnarray}}

\def\ee{\end{eqnarray}}

\def\ba{\begin{array}}

\def\ea{\end{array}}

\begin{document}

\preprint{1}

\title{Semi-classical theory of Aharonov-Bohm oscillations under quantized Hall conditions}

\author{A. S\i dd\i ki} %
\affiliation{Department of Physics, Faculty of Sciences, Istanbul University, 34134 Vezneciler-Istanbul, Turkey } \affiliation{Department of Physics, Harvard University, Cambridge, 02138 MA, USA}
\author{K. G\"uven}
\affiliation{Department of Physics, Ko\c c University, 34450 Istanbul, Turkey}
\begin{abstract}
We demonstrate that the experimentally observed conductance oscillations through a gate-controlled quantum dot under quantized Hall conditions originate from a true Aharonov-Bohm phase that is acquired by the electrons as they co-propagate coherently through incompressible strips at a particular filling factor and enclose a certain magnetic flux within the dot. The absence of oscillations at certain filling factors is explained. The self-consistent screening theory of the integer quantized hall effect reveals the range of experimental parameters under which further intriguing oscillation patterns are observable.

\end{abstract}




\maketitle

\begin{figure}[t] {\centering
\includegraphics[width=.6\linewidth]{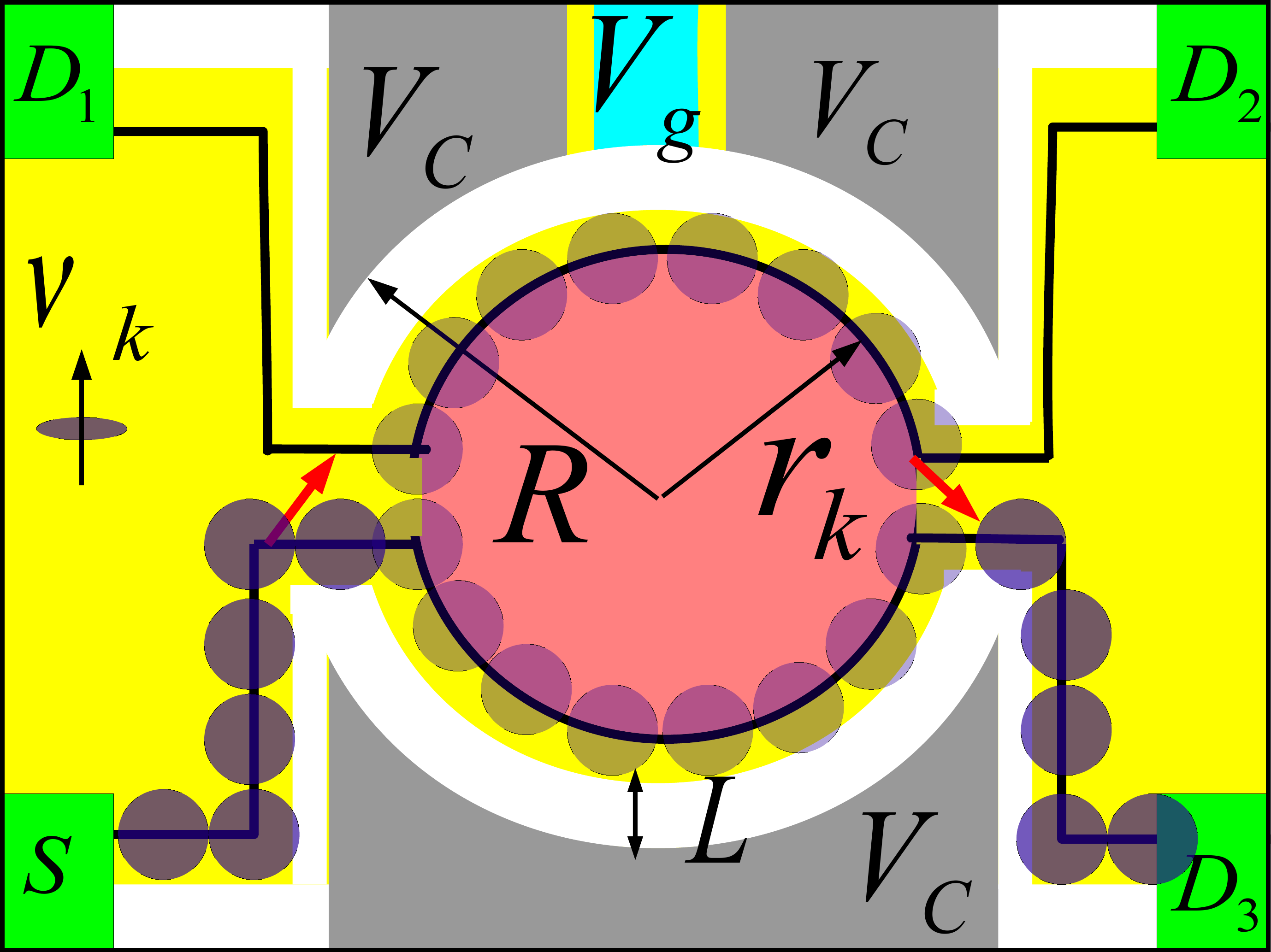}\vspace{0mm}
\caption{\label{fig1}(Color online) Generic sample geometry. The circles depict the coherent superposition of cyclotron orbits carrying an intrinsic phase ($\propto \nu_k$), and enclosing a certain area of radius $r_k$ within the quantum dot giving rise to a phase proportional to $N_k$ (pink). Boxes at the corners are the contacts.}}
\end{figure}

A gate-controlled quantum dot constriction in a two-dimensional electron system (2DES) under quantized Hall conditions exhibit unusual transport properties yet to be explained. One feature is the oscillation period ($\Delta B$) of the conductance, ($\delta G$), through the quantum dot by the varying gate potential ($\Delta V_g$), magnetic field ($B$) or the constriction size, where a seemingly controversial or sample-specific dependence is observed. Quite a number of experimental ~\cite{Heiblum03:415,Goldman05:155313,goldman07:e/3,godfrey:07,Marcus09,Bernd:ABosc.,goldman:interactions,Nissim09:} and theoretical~\cite{bernd:07,igorAB08,Engin:09japon,Esa:ab,Kotimaki:10,Halperin:10} efforts were devoted to exploit these quantum oscillations in analogy to the Aharonov-Bohm (AB) effect in semiclassical magneto-transport. A typical experimental sample geometry is as shown in Fig.~\ref{fig1}. The 2DES resides in the yellow plane, the dark regions are the gated areas defining the quantum dot with radius $R$, area $A$. Current flows through the two openings. A side gate $V_g$ (blue) regulates the area of the interferometer and the path of the current channel(s). $L$ is the depletion length. In the following, the reported experimental observations are summarized:
\begin{itemize}
\item For the trench-gated samples (where the edges are first etched then metal is deposited ~\cite{Goldman05:155313,godfrey:07,goldman:interactions}): \emph{(i)} Oscillation period $\Delta B$ is almost linear throughout the inter-plateau interval with weak dependence on $B$. \emph{(ii)} No pronounced interference oscillations are observed between odd and even integer plateaus (except the first plateau). \emph{(iv)} $\delta G$ has positive slope in the $B-V_g$ plane.
\item For the gated samples: \emph{(i)} At small dots ($A\approx2~\mu$m$^2$) the observations are similar to that of trench-gated samples. \emph{(ii)} $\Delta B$ is independent of $B$ when no channel is fully transmitted.~\cite{Nissim09:} \emph{(iii)} At large dots ($A\approx18~\mu$m$^2$), the slope of $\delta G$ is negative, contrary to trench-gated samples.~\cite{Marcus09}
\end{itemize}

One route to explain these observations follow the Landauer-B\"uttiker formalism through edge channels (LBEC) formed due to the physical edges of the constriction. In contrast to the Aharonov-Bohm effect, however, these channels are counter-propagating along the opposing edges of the system; partitioning takes place via backscattering~\cite{Halperin:10}.

In this work, we take the path less traveled by, through the self-consistent screening theory of integer quantum Hall effect,~\cite{siddiki2004} to explain the underlying mechanism of conductance oscillations. In a nutshell: (1) The theory calculates the electron-density profile analytically by including electron interactions at Hartree level and determines the current path. (2) When the path is narrow enough to accommodate only one electron wavefunction of finite extend, the current is coherent. (3) The co-propagating paths accumulate phase as a function of control parameters and interfere at the other end, leading to the observed oscillations. We demonstrate that this theory is capable of addressing all the reported experimental results, almost quantitatively.

We begin by describing the analytical model which determines the density profile and the incompressible strips (ISs), within the Thomas-Fermi (TFA) - and quasi-Hartree-approximations (QHA). Based on the model, the conditions to observe current interference under quantized Hall regime are prescribed. We then analyze the experimental results by virtue of the model and investigate further intriguing oscillation patterns.
%
%

%

\emph{The model}: The single particle Hamiltonian in the presence of normal magnetic field, $B$, and the electrostatic potential, $V(r)$, is given by
$H^{s}=\frac{1}{2m^*}(\textbf{p}(r)-\frac{e}{c}\textbf{A}(r,z))^2+\hat{s}g^*\mu_BB+V(r),$ where $m^*$  is the effective mass and $\textbf{p}(r)$ is the momentum of the particle. The second term accounts for the Zeeman splitting given by the effective $g^*$ factor, spin $\hat{s}=\pm1/2$, and the Bohr magneton $\mu_B$. $V(r)$ consists of the confinement $V_{\rm conf}(r)$ and Hartree $V_H(r)$ potentials within mean-field approximation.

\emph{Thomas-Fermi approximation}: Using the Landau gauge, Hartree approximation yields the Landau energies $E_n(X_0)=E_n+V(X_0)$. Here, $X_0=-l_b^2k_y$ is the center coordinate, $k_y$ is the quasi-continuous momentum, $l_b=\sqrt{1/(2\pi \Phi_0 B)}$ ($\Phi_0=h/|e|$, $e<0$ is the electron charge and $h$ stands for Planck constant) is the magnetic length and $E_n=\hbar\omega_c(n+1/2)$, with the cyclotron frequency $\omega_c=eB/m^*$. The spin-resolved Landau wavefunctions (LWs) are
$\psi^{s}_{n,X_0}(x,y)=Me^{ik_yy}e^{-(\xi)^2}H_n(\xi), \label{eq:wavefunction}$ where $H_n(\xi)$ are the Hermite polynomials of order $n$ ($\xi=(x-X_0)/2l_b$), with a normalization factor $M$. Assuming that the potentials vary slowly on the quantum mechanical length scales, TFA, one writes the LWs as $|\psi^{s}_{n,X_0}(x,y)|^2=\delta(x-X_0)$ and energies as $E_n(X_0)=E_n+V(x)$. Later, the effects of many-body interactions are included through the Thomas-Fermi-Dirac approximation~\cite{GonulTFD:09}. We note that the LWs are the \emph{coherent} superpositions of the classical cyclotron orbits with the same center coordinate $X_0$ and different $Y$, provided that the extend of the wavefunction ($\sim l_b$) is less than the classical orbit. Remarkably, this leads to an \emph{intrinsic} AB phase which manifests the number of electrons orbiting a single flux (dashed-circles in Fig.~\ref{fig1}) i.e. the filling factor $\nu$ of the Landau levels. For example, if two electrons orbit one flux quantum ($\nu=2$), the AB oscillation period will be twice that of a single electron. This intrinsic phase is quantized at the current paths and will determine the overall phase together with the area enclosed by the paths (pink area in Fig.~\ref{fig1}) that the particle travels.

We prescribe the electrostatic density profile of the electrons within the quantum dot by the function $ n(r)=n_0(1-e^{-(|R-L|-r)/t})$ if  $r<|R-L|,$ and being zero otherwise. Here, $n_0~(=3\times10^{11}$ cm$^{-2})$ is the bulk electron density, $R~(=120)$ is the dot radius, $L~(=20)$ is the depletion length, and $t$ is the width of the \emph{electron poor} region following the depletion, measured in units of effective Bohr radius $a_B^*=9.81$ nm. Our rationale for this functional form stems from a modified version of the non-self-consistent TFA theory of Chklovskii \emph{et al.}~\cite{Chklovskii92:4026} taking into account the fact that the actual density profile close to the edges are quite different~\cite{Halperin94:etchedge,Lier94:7757}. A similar discussion of the electrostatics is provided in~\cite{Goldman05:155313}, using another modified Chklovskii density distribution without spin and assuming $R\gg L$, where the dot area $A$ can be taken constant. We set $t\gtrsim3$ for the gated samples. This set of parameters approximates the self-consistent calculation of density distributions very well~\cite{Lier94:7757}. The spatial positions of the integer filling factors $\nu(r_k)=k=2\pi l_b^2 n(r_k)$, are obtained by $r_{k}=R+t\ln(1-k/\nu_0),$ where $\nu_0$ is the bulk filling factor and $k=1,2,3\cdot\cdot\cdot$ an integer, provided that $\nu_0>k$. Following Ref.~\cite{Chklovskii92:4026}, the width of the ISs, within the TFA is given by $a^{\rm TFA}_k=[(\frac{4\alpha a_B^*}{\pi\nu_0})t]^{1/2}\exp{[(\frac{|R-L|-r_k}{2t})]}$, where $\alpha$ is the energy gap between the quantized levels. We describe the odd integer gaps by $\alpha~(=\frac{g^*\mu_BB}{\hbar\omega_c})$ and even gaps by $1-\alpha$. To this end, the electrostatic model above is useful in estimating $r_k$ and $a_k$. Self-consistent numerical calculations provide more precise results~\cite{siddiki2004}.
\begin{figure} {\centering
\includegraphics[width=1\columnwidth]{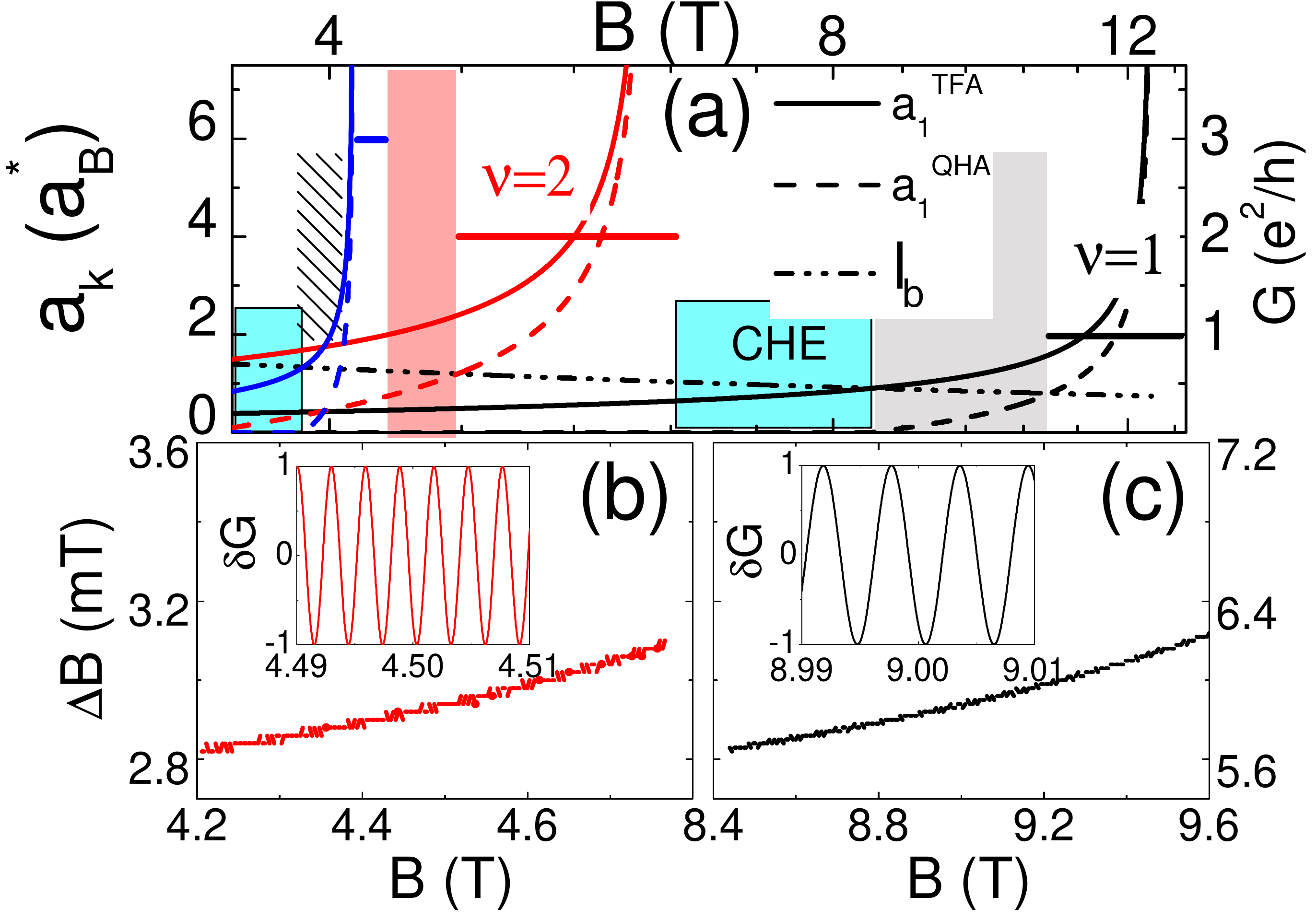}\vspace{-0cm}
\caption{\label{fig2} (a) The widths of the ISs as a function of $B$. The (color) shaded areas indicate the interference interval, whereas diagonal pattern shows the overlap interval. Horizontal thick lines denote the plateaus. The conductance fluctuations (inset) and $B$ periodicity obtained considering (b) $\nu=2$ and (c) $\nu=1$. Notice the twice periodicity of $\nu=2$ compared to $\nu=1$. Here $t=3$.}}
\end{figure}

\emph{Finite extent of the wavefunctions:} The drawback of the TFA theory is the underestimation of the finite extent of the LWs that ultimately violates the assumption of slowly varying potential at narrow ISs. This is cured by incorporating the finite width of the LWs within the QHA, where the wavefunction assumes the form $\psi^{s}_{n,X_0}(x,y)\approx\psi^{s}_{n,0}(x,y)$, whereas the eigen-energies are still given as in TFA. Such an improvement is shown to be essential~\cite{siddiki2004} for the IQHE. The IS widths can be approximated as $a_k^{\rm QHA}=(1-\frac{l_b}{a_k^{\rm TFA}})a_k^{\rm TFA} \label{eq:akQHA}$, within the QHA.

\emph{Interference regime under quantized Hall conditions:} Equipped with the improved density profile and wavefunctions, we now calculate the widths of the ISs and the number of magnetic flux quanta $N_k$ enclosed by the IS at filling factor $\nu=k$. Fig.~\ref{fig2}a, shows the calculated $a_k^{\rm TFA}$ (solid lines) and $a_k^{\rm QHA}$ (broken lines) as a function of $B$. Since interferometers are defined on high mobility structures, we incorporate the exchange enhancement of $g^*$ by setting $\alpha=0.2$. We look for the interference conditions by determining the plateau interval at first, considering $\nu=1$. It is shown that, if $a_k^{\rm QHA}> l_b$ the incompressible strip decouples the edges, (back)scattering is suppressed and the plateau sets in~\cite{siddiki2004}, $B>10.5$ T. Within the Drude model, the longitudinal conductivity $\sigma_L$ vanishes. Thus, all the current is confined to the strip, and the Hall conductivity $\sigma_H=\nu_k e^2/h$, is quantized. But the current is not coherent since the IS can accommodate multiple electrons. In the case of $a_k^{\rm TFA}< l_b$, the cyclotron motion is no longer quantized and we observe the classical Hall effect (CHE), $B<8.4$ T. Hence, we are left with the interval $a_k^{\rm TFA}> l_b \gtrsim a_k^{\rm QHA}$, where only \emph{one coherent state resides inside the strip}. The definite meaning of incompressibility is blurred in this case, since electrons can tunnel across the strip. Therefore, the name \emph{evanescent incompressible strip} (eIS) is introduced~\cite{Sailer:10}. We argue that the AB oscillations are observable as long as the varying $B$ maintain the interval condition above, since the strips preserve the phase coherence and interfere at the edges through tunneling. For $\nu=2$, the pattern is very similar.

Our calculations show that the $\nu=3$ plateau sets in before $a_2^{\rm TFA}< l_b$ is reached and thus two eISs survive in the expected interval of interference for $\nu=3$. This smears out the interference at the output terminal, since now the electron can tunnel between the two eISs, since $|x_3-x_2|\approx l_b$. Therefore, the area $A_3$ cannot be determined definitely. This remarkable behavior agrees well with the experimental findings such that the AB interference cannot be identified at odd integer filling factors mostly (except $\nu=1$), and the interference at $\nu=3$ is only once reported at the etched sample. We observe interference at $\nu=3$ for etched samples by setting $t\sim 1$ (not shown here).
\begin{figure} {\centering
\includegraphics[width=0.90\columnwidth]{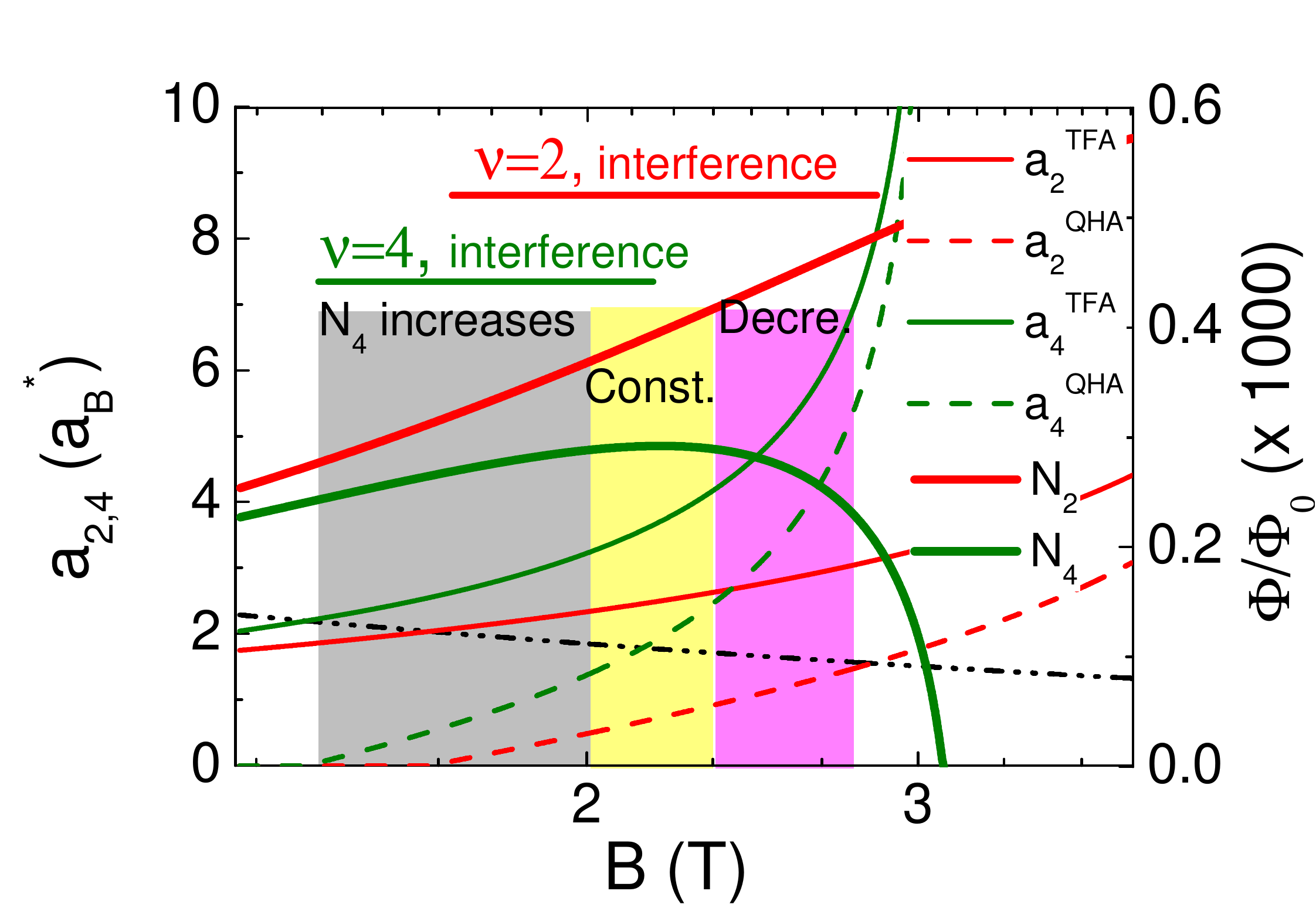}
\caption{\label{fig3}The variation of flux number encircled by $\nu=2$ (red) and $\nu=4$ (dark green) evanescent strips. In the interval $[2,2.33]$, $N_4$ stays constant within line thickness.}}
\end{figure}
\begin{figure*}[ht] {\centering
\includegraphics[width=0.53\columnwidth]{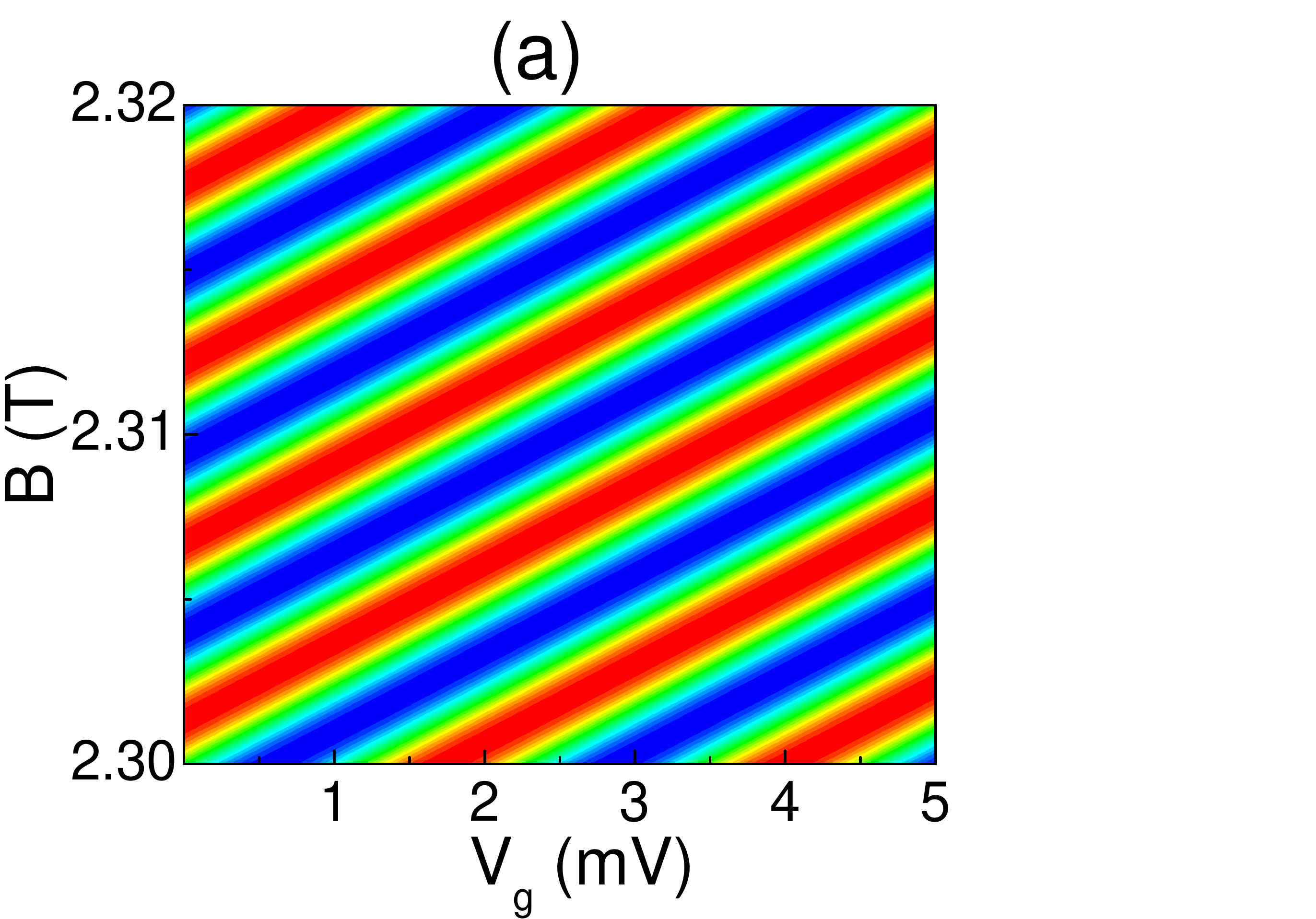}\hspace{-1.35cm}
\includegraphics[width=0.53\columnwidth]{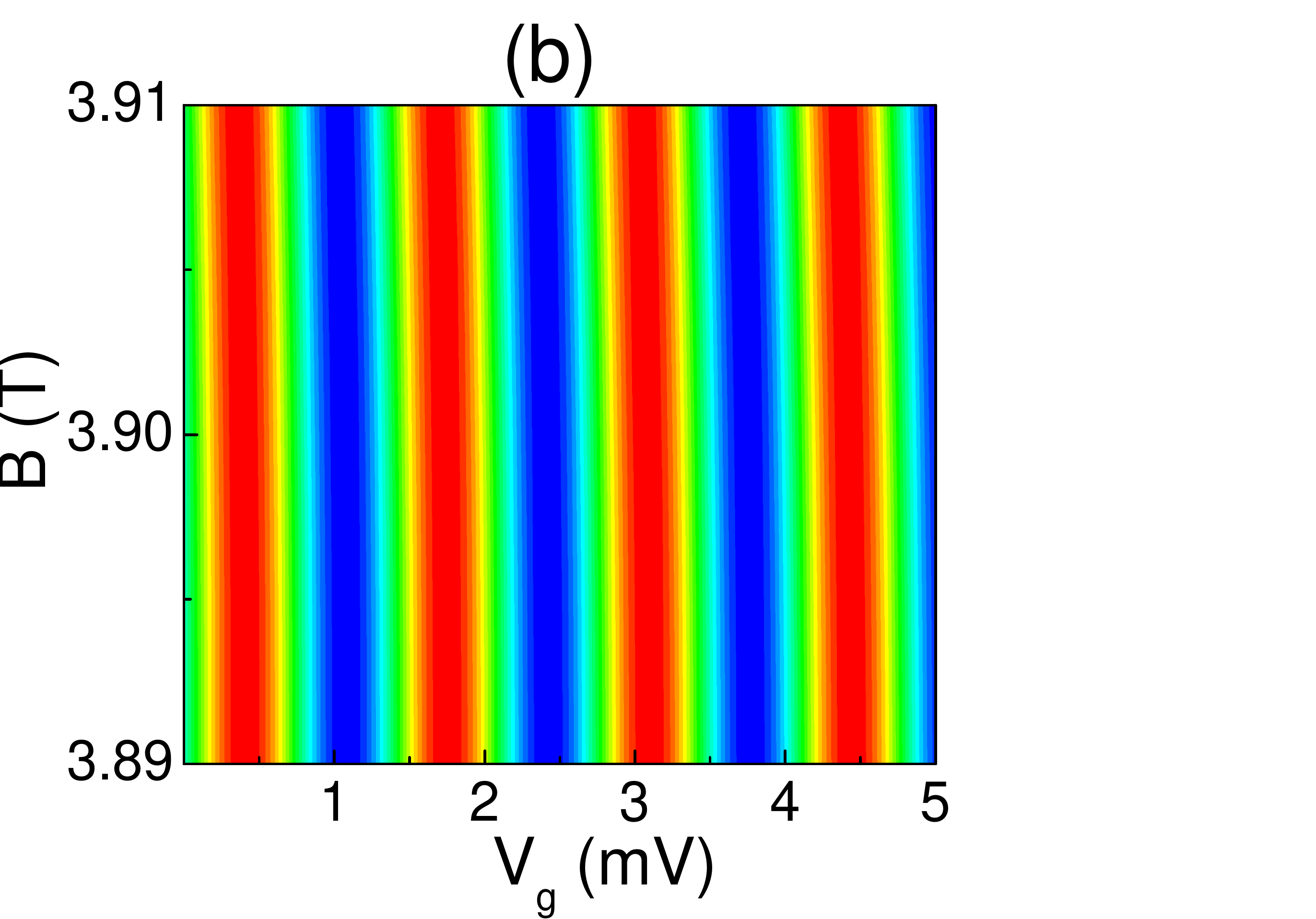}\hspace{-1.35cm}
\includegraphics[width=0.53\columnwidth]{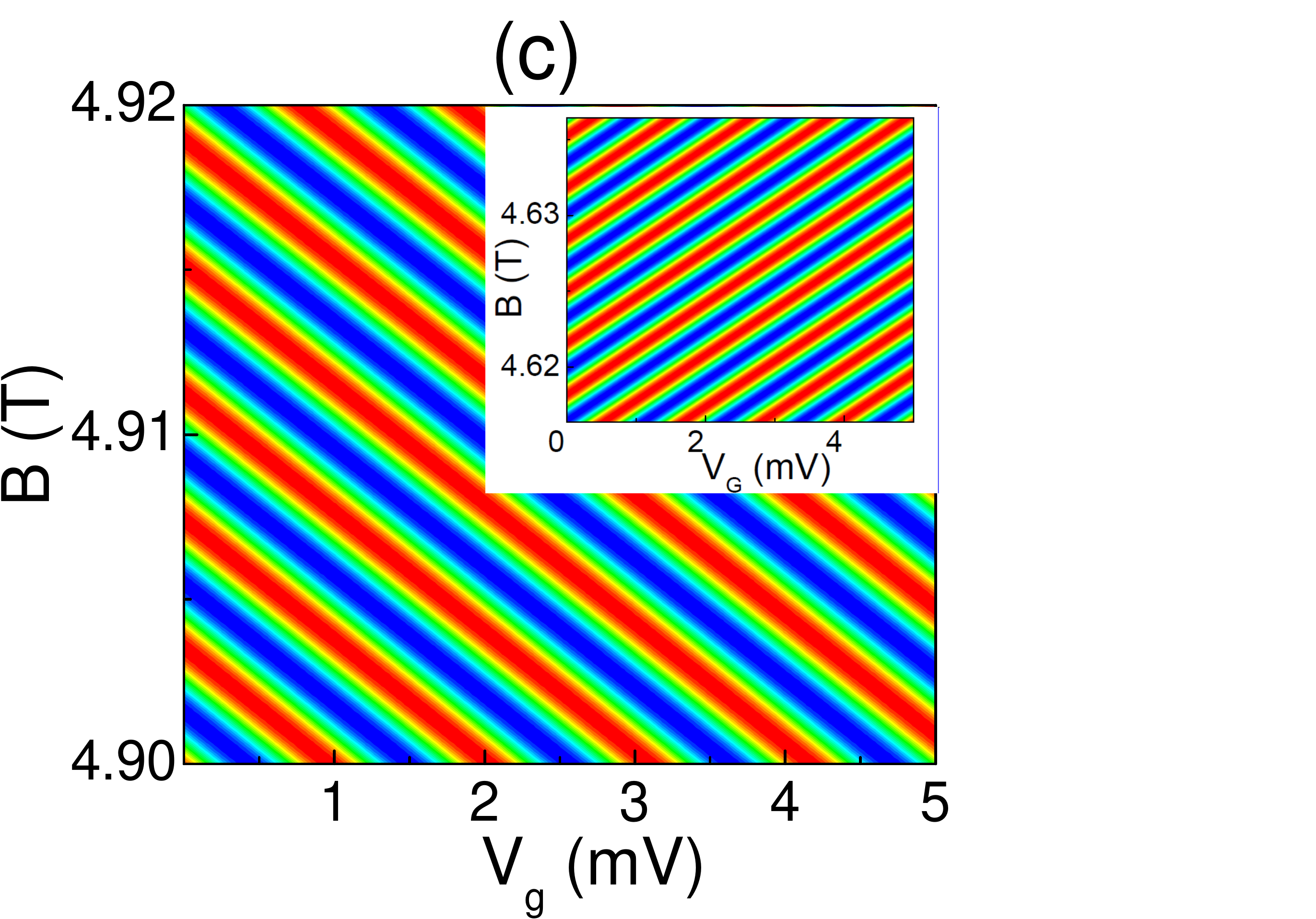}\hspace{-1.35cm}
\includegraphics[width=0.53\columnwidth]{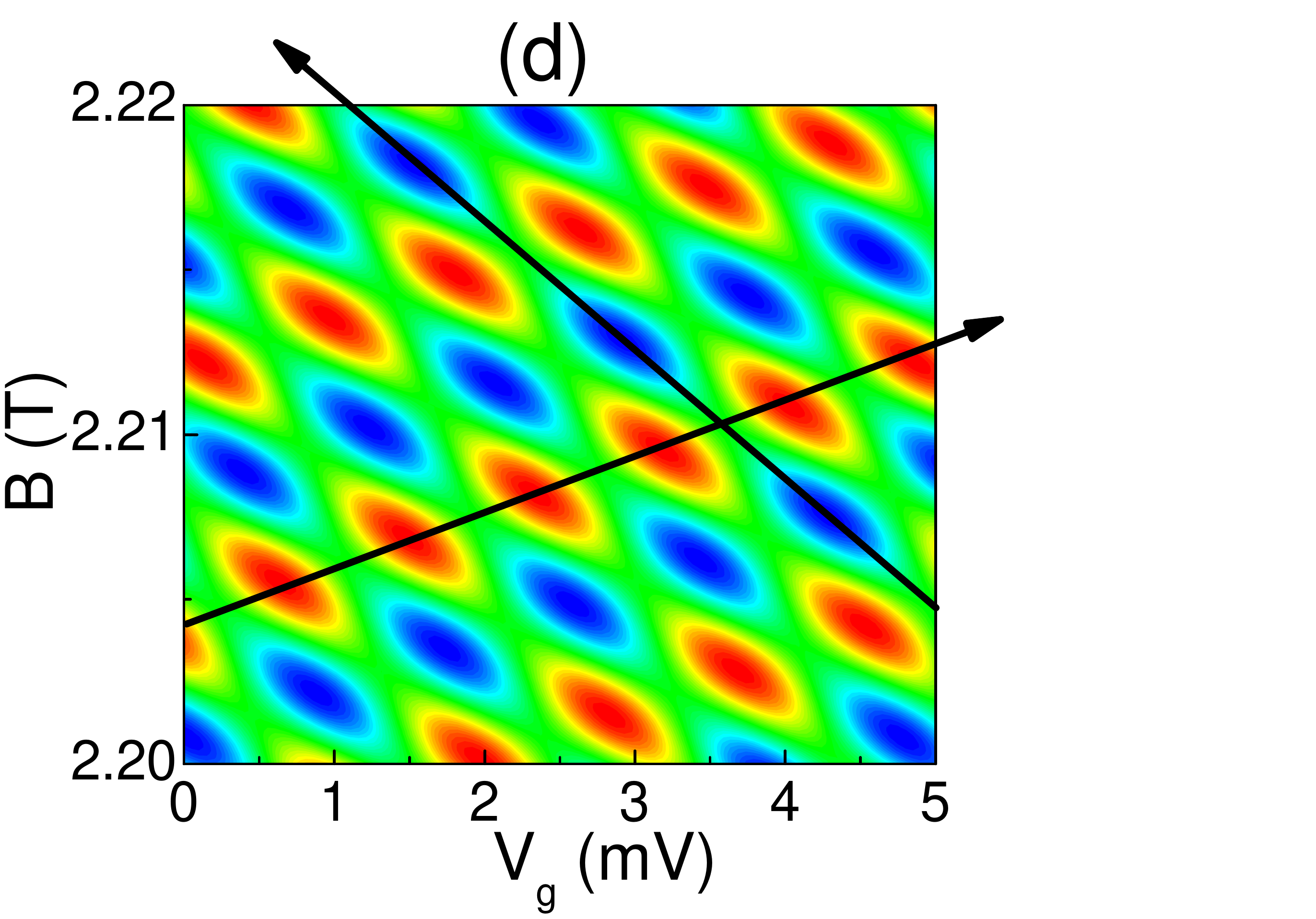}\hspace{-1.35cm}
\includegraphics[width=0.53\columnwidth]{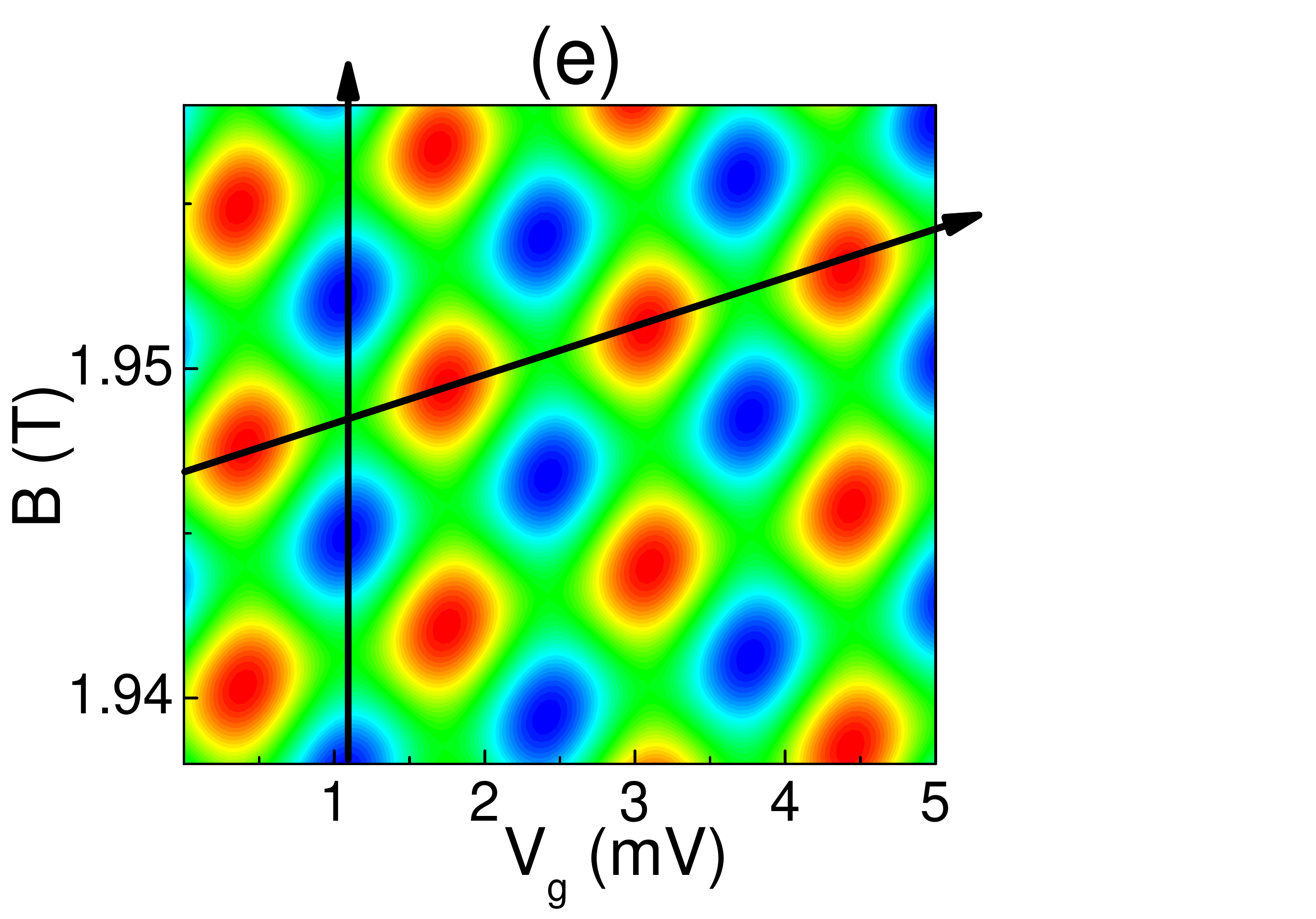}
\caption{\label{fig4} The calculated interference patterns for $\nu=2$ plateau with $t=20$ and $R=120$. (a) $N_2$ increases linearly with $B$ (b) is constant (c) decreases. The superposition of interference patterns $\nu=1$ and $\nu=4$ states, where $N_1$ increases linearly, meanwhile $N_4$ decreases (d) or is constant (e). The inset in (c) is the corresponding calculation for a steep edge defined sample, $t=3$ and shows an opposite slope compared to smooth edge sample. The gate potential periodicity $\Delta V$ depends strongly on the length of the plunger gate and the shown $V_g$ are relative to 500 mV.}}
\end{figure*}
In Fig.~\ref{fig2}b-c the $B$ periodicity is shown, together with the calculated conductance oscillations $\delta G = G_0 e^{2\pi N_k\nu_k}$ as insets, with $G_0=e^2/h$. Here, $N_k$ varies almost linearly with $B$. Clear oscillations at $\nu=2$ (b), $\nu=1$ (c) are seen and their periods are determined by the $\nu_k$. This becomes more clear if one plots the period of the oscillations $\Delta B$ as a function of $B$, where approximately linear variation is observed for regular oscillations imposed by the areal dependency on $B$ of the interfering channels. Moreover, our model predicts a linear variation of $\sim \%10$ that is reported at the experiments and is attributed to the weak dependence of the interferometer to the enclosed area. Note that, $N_k$ determines whether the particle will end in $D_2$ or $D_3$~\cite{Kotimaki:10}.

The variation of $N_{2,4}$ is shown in Fig.~\ref{fig3} as a function of $B$, where a smoother edge profile is imposed ($t=7$). We see that $N_4$ can increase ($B<2$ T), stay constant ($2<B<2.33$ T) or decrease while increasing $B$, since the area enclosed $A_k=\pi r_k^2$ also depends on $B$, if $R$ is not considerably larger than $L$. Such an areal dependency imposes two interesting cases i) if $N_k$ stays constant while increasing $B$, $\delta G$ becomes independent of $B$, hence, $\Delta B$ ii) if the area shrinks faster than the increase of $B$, then $\Delta B$ decreases with increasing $B$. A constant-remaining $A_k$ (i.e. linear increase of $N_k$) is assured in two cases; i) by steep edge definition $t\lesssim3$ and/or ii) by producing very large devices, that is $R\gg L$. In such devices $\Delta B$ is only determined by the filling factor of interfering eIS, referred as the AB dominated regime in Ref.~\cite{Halperin:10}.

The effect of a plunger gate on $\delta G$ is somewhat straightforward: The enclosed area is reduced by an amount directly proportional to the length of the gate and $V_g$, hence if one flux quantum is excluded the conductance shifts a whole period. We stress that there is no Coulomb blockade (CB), since the system is \emph{completely} compressible, within the interference interval. We show the calculated interference patterns in Fig.~\ref{fig4}(a-c) considering a gated sample at the $\nu=$2-3 transition. In Fig.~\ref{fig4}a $a_2^{\rm TFA}>l_b>a_2^{\rm QHA}$ condition is satisfied, $N_2$ increases linearly with $B$ and $\nu=2$ eIS carries the current, whereas in Fig.~\ref{fig4}b $N_2$ stays constant within a percent of a flux quanta. Such a small variation, however, will be smeared out if a full self-consistent calculation is performed, together with charging effects. Fig.~\ref{fig4}c, shows the different oscillation slopes considering a gated or etched (inset) samples. At etched samples, the interference area is almost independent of the $B$ field due to steep density profile at the edges and presents similar behavior with gated samples, where $N_k$ increases linearly with $B$. In Fig.~\ref{fig4}(d), two states with $\nu=1$ and $\nu=4$ are partially transmitted, here $N_1$ increases and $N_4$ decreases. Note that, the two current channels are far apart (i.e. $|x_4-x_1|\gg l_b$), hence, the overlap is not significant as it was among $\nu=2$ and $\nu=3$ discussed previously. The $V_g$ dependency remains roughly unaffected. All the experimental patterns reported in Ref.s~\cite{Nissim09:,Marcus09} can be obtained by the areal dependency of the interfering channel without making \emph{ad hoc} assumptions. Therefore, we argue that the observed oscillations are due to genuine AB phase. An important issue is that the case in Fig.~\ref{fig4}c is recovered for the case Fig.~\ref{fig4}a, if a top gate is negatively biased: It pushes electrons towards edges and the density profile becomes steeper similar to an etched sample. At gated large samples, the electron density becomes constant within the dot, hence the smooth variation of the density is removed, resulting in increase of $N_k$ with increasing $B$. Indeed, a self-consistent investigation of the effect of a top gate, supports our above arguments~\cite{Lier94:7757}. The depicted interference patterns are obtained by virtue of the effect of steepness of the density profile on the interference area only: The variation of the oscillation period is directly linked to the enclosed area, which, in turn, depends on the density profile.

\emph{Remarks:} \emph{i)} The eISs carry the \emph{co-propagating} dissipative current. However, dissipation is strongly suppressed at extremely low temperatures, namely $\sigma_L$ becomes exponentially small and backscattering is suppressed, whereas $\sigma_H$ is quantized at \emph{T=0}. This also explains why the experiments have to be performed at mK, although the energy gap is at the order of Kelvins. At these temperatures, the semi-classical description of the transport resembles the quantum mechanical one. \emph{ii)} The areal independence of the $\delta G$ at other particle interferometers is due to etched samples, and is consistent with our findings. \emph{iii)} Our theory is consistent with Goldman group's interpretation~\cite{goldman:interactions}. \emph{iv)} Inclusion of charging effects to our model, modifies $\Delta (V,B)$ at the level of few percent. Since, the phase period is determined by $\nu_k$.

\emph{Summary:} Employing the screening theory of the IQHE to the AB effect observed at 2DES, we showed that: \emph{i)} The $\nu=3,5..$ oscillations are commonly suppressed due to the finite overlap between eISs with different $\nu$. \emph{ii)} The weak linear $B$ dependency of the period is explained by the non-linear variation of the interference area \emph{iii)} The reported oscillations are due to genuine quantum mechanical phase of co-propagating coherent electrons. \emph{iv)} The different interference patterns observed are by the virtue of non-trivial $N_k$-$B$ dependence on enclosed area, via strong influence of the density profile due to device properties such as edge definition and constriction size.

\end{document}